\documentclass[preprint]{revtex4-1}

\usepackage{amsmath}
\usepackage{graphicx}

\newcommand{\qvec}{{\bf q}}
\newcommand{\pvec}{{\bf p}}
\newcommand{\mvec}{{\bf m}}
\newcommand{\goo}{g_{\mbox{\tiny OO}}}

\begin{document}

\title{{\it Ab initio} mass tensor molecular dynamics}

\author{Eiji  Tsuchida \\ 
   Nanosystem Research Institute, \\
   National Institute of Advanced Industrial Science and Technology (AIST), \\
   Tsukuba Central 2, Umezono 1-1-1, Tsukuba 305-8568, Japan }

\begin{abstract}
Mass tensor molecular dynamics was first introduced by Bennett 
[J.~Comput.~Phys. {\bf 19}, 267 (1975)] for efficient sampling 
of phase space through the use of generalized atomic masses. 
Here, we show how to apply this method to 
{\it ab initio} molecular dynamics simulations 
with minimal computational overhead. 
Test calculations on liquid water show a threefold reduction 
in computational effort 
without making the fixed geometry approximation. 
We also present a simple recipe for estimating the optimal atomic masses using only 
the first derivatives of the potential energy. 
\end{abstract}

\maketitle

\newpage

\section{Introduction}

The efficiency of molecular dynamics (MD) simulations is 
governed by the choice of time step, 
which is usually taken 1/15 - 1/20 
of the period of the fastest oscillation in the system. 
In liquid water, for instance, the O-H stretching mode has 
a period of 9-10 fs, thus limiting the time step to less than 1 fs. 
On the other hand, the length of a simulation must be 
much longer than structural relaxation times of the system, 
which range from picoseconds for liquid water to 
microseconds for biological systems \cite{KLSH}. 
Therefore, much effort has been made to 
reduce this gap in time scale \cite{CLREV1,CLREV2}, 
e.g., by eliminating the high frequency motion, 
or equivalently, by accelerating the low frequency motion. 

One of the most widely used technique in classical MD is 
constrained molecular dynamics (CMD) \cite{SHK,RAT}, in which 
the bond lengths and angles may be fixed to their equilibrium values. 
CMD allows us to increase the time steps by a factor of two to three, 
due to the absence of fast intramolecular vibrations. 
CMD is equally valid in {\it ab initio} 
molecular dynamics (AIMD) simulations \cite{AIRAT,CPRAT} 
which provide more accurate interatomic forces 
at the expense of much higher computational costs \cite{CP,DFTREV2,DFTREV3}. 
We can also use CMD to prevent (undesirable) breaking 
of covalent bonds in the initial stages of the simulations \cite{SPES}. 
However, bond flexibility can play an important role 
if two or more phases coexist in the simulation cell \cite{CHVG,RASA}. 
Furthermore, overuse of bond angle constraints is prone to 
numerical artifacts \cite{CLREV1}. 

A more recent development, which is free from these drawbacks, 
is the multiple time step (MTS) algorithms like Verlet-I \cite{GRUB} 
and r-RESPA \cite{RESPA}. 
In MTS, one can use large time steps for computationally 
expensive long-range interactions, 
while inexpensive short-range ones 
are integrated with small time steps. 
MTS is superior to CMD in that no fixed geometry approximation 
is required, while a similar or higher gain in performance can be achieved. 
Unfortunately, interatomic forces from {\it ab initio} calculations 
cannot be divided into short- and long-range components exactly. 
Therefore, with the exception of approximate implementations \cite{MIXFOR1,MIXFOR2}, 
MTS is not generally used for AIMD. 

In the present paper, we investigate an alternative approach 
which takes advantage of the extra degrees of freedom 
associated with the choice of atomic masses. 
A simple approach along these lines is to rescale 
the atomic masses appropriately, which is particularly effective 
in systems containing hydrogen \cite{POMc,FEEN,ZWZ,WW}, 
as well as for adiabatic free energy calculations \cite{RMZT,CAFES}. 
It is also a common practice to replace hydrogen by 
deuterium in AIMD study of liquid water \cite{WAT2}, 
thereby increasing the time step by a factor of $\sqrt{2}$. 

Whereas the mass scaling method alleviates the time-scale problem to some extent, 
it does not fully eliminate the strong anisotropy of 
the phase space caused by covalent bonds. 
This problem can be overcome by using a more general formulation 
termed mass tensor molecular dynamics (MTMD), 
which was first introduced by Bennett more than three decades ago \cite{MTMD}. 
MTMD enables us to make the phase space nearly isotropic 
by using a nondiagonal position-dependent mass tensor, 
thus leading to enhanced sampling. 
Bennett used a constant, but nondiagonal mass tensor 
in his calculations on a simple polymer chain, 
and achieved a five- to tenfold reduction in computational effort. 
The staging method used in path integral MD \cite{DFTREV2} can also 
be viewed as a variant of MTMD. 
Melchionna has recently developed a similar method 
using internal coordinates in a series of papers \cite{MEL1,MEL2,MEL3}. 
In the present work, we show how to apply MTMD to AIMD simulations 
with minimal computational overhead. 
We also demonstrate the effectiveness of our approach 
in Born-Oppenheimer MD simulation of liquid water.

\section{Theory}
The classical Hamiltonian for a system of $N$ atoms is given by 
\begin{equation}
\label{OHAM}
H_0 (\qvec,\pvec) = \frac12 \sum_{i} \frac{p_i^2}{m_i} + U(\qvec), 
\end{equation}
where $\mvec$, $\qvec$ and $\pvec$ are vectors of dimension $3N$, 
representing atomic masses, positions and momenta. 
$U(\qvec)$ denotes either the classical potential energy function \cite{FFREV1}, 
or the Kohn-Sham total energy \cite{DFTREV1,DFTREV4}. 
In what follows, we will focus on the latter problem. 

The generalized Hamiltonian introduced by Bennett is given by \cite{MTMD} 
\begin{equation}
\label{GHAM}
  H (\qvec,\pvec) = \frac12 \sum_{i,j} p_i \, M^{-1}_{ij} (\qvec) \, p_j + U(\qvec), 
\end{equation}
where $M$ is a positive definite, symmetric matrix of dimension $3N$ 
which depends on the atomic positions \qvec. 
Let us assume that $\rho(\qvec)$ is the canonical probability distribution 
in configuration space: 
\begin{equation}
\label{PROB}
\rho(\qvec) = \frac{\int d\pvec \, \exp (- \beta H)}{\int d\qvec \, d\pvec \, \exp (-\beta H)}, 
\end{equation}
where $\beta = 1 / k_{\rm B} T$ is the inverse temperature. 
Then, the ensemble average of any operator $A(\qvec)$ is given by 
\begin{equation}
\left< \, A \, \right> = \int A(\qvec) \rho(\qvec) d\qvec. 
\end{equation}
If Eq.(\ref{GHAM}) is substituted into Eq.(\ref{PROB}), 
and integration over $\pvec$ is performed \cite{IMA}, we obtain 
\begin{equation}
\rho(\qvec) = \frac{|M (\qvec)|^{\frac12} \exp (- \beta U (\qvec))}
{\int d\qvec \, |M (\qvec)|^{\frac12} \exp (- \beta U (\qvec))}. 
\end{equation}
Therefore, as long as $\det(M) = |M (\qvec)|$ is independent of $\qvec$, 
$\rho(\qvec)$ (and thus $\left<\,A\,\right>$) agrees with that of 
the original Hamiltonian. 
The same conclusion holds true for the microcanonical ensemble \cite{MTMD}. 

There are numerous ways to choose $M$ 
that satisfies the above conditions, 
but a reasonable choice is 
\begin{equation}
\label{MHEQ}
M \approx {\cal H} \times c (\qvec), 
\end{equation}
where the Hessian matrix ${\cal H}$ is given by 
\begin{equation}
{\cal H}_{ij} = \frac{\partial^2 U (\qvec)}{\partial q_i \partial q_j},  
\end{equation}
and $c (\qvec)$ is a scalar function to compensate for 
the $\qvec$-dependence of $|{\cal H}|$. 
If $U(\qvec)$ is a convex quadratic function of $\qvec$, 
Eq.(\ref{MHEQ}) is optimal in the sense that 
all normal modes have the same frequency. 
In practice, it is prohibitively expensive to calculate the exact values of 
${\cal H}_{ij}$ at each step of AIMD. 
Instead, we introduce a simple harmonic approximation to $U(\qvec)$, given by  
\begin{equation}
\label{VHARM}
V_{\mbox{\scriptsize H}} (\qvec) = 
\sum_{\rm angles} C_a (\theta - \theta_{\rm eq})^2 + 
\sum_{\rm bonds} C_b (r - r_{\rm eq})^2,   
\end{equation}
where only covalent bonds contribute to the sum. 
Then, we can define 
\begin{equation}
V_{ij} = \frac{\partial^2 V_{\mbox{\scriptsize H}} (\qvec)}{\partial q_i \partial q_j}, 
\end{equation}
which is a sparse, symmetric, and indefinite matrix. 
The simplest way to construct a positive definite matrix 
similar to $V$ is to define 
\begin{equation}
\label{NAIVE}
M = (V+ \epsilon I) \times c (\qvec)
\end{equation}
using a small positive constant $\epsilon$. 
In our experience, however, the value of $\epsilon$ tends to be relatively large 
in finite temperature simulations, which has a negative impact on 
the performance of MTMD. 
Therefore, we adopt a more robust definition 
based on the eigenvalue decomposition of $V$ \cite{MTMD}. 
$V$ is first diagonalized by an orthogonal matrix $P$ as 
\begin{equation}
V=P^T \cdot \Lambda \cdot P, 
\end{equation}
where $\Lambda$ is a diagonal matrix given by 
$\Lambda_{ij} = \lambda_i \delta_{ij}$. 
Then, we introduce 
\begin{equation}
M_0=P^T \cdot f(\Lambda) \cdot P, 
\end{equation}
where $f(\lambda)$ is a filtering function given by 
\begin{equation}
\label{FILTER}
f(\lambda)=(\lambda^6+\epsilon^6)^{\frac16}. 
\end{equation}
At variance with Eq.(\ref{NAIVE}), $M_0$ is a positive definite matrix 
for any nonzero $\epsilon$. 
Finally, $M$ is given by 
\begin{equation}
M = c_0 \, M_0, 
\end{equation}
where 
\begin{equation}
c_0 = \left( \frac{\alpha}{|M_0|} \right)^{\frac{1}{3N}},  
\end{equation}
and $\alpha$ is an arbitrary normalization factor 
which does not depend on $\qvec$. 
By definition, $\det(M)= \alpha$ holds for any $\qvec$, 
irrespective of the choice of $V_{\mbox{\scriptsize H}}$ and $\epsilon$. 

The equations of motion derived from 
the generalized Hamiltonian of Eq.(\ref{GHAM}) are given by 
\begin{equation}
\label{NT1}
\frac{dq_k}{dt} = \frac{\partial H}{\partial p_k} = \sum_j p_j \, M^{-1}_{kj} (\qvec)
\end{equation}
and 
\begin{equation}
\label{NT2}
\frac{dp_k}{dt} = -\frac{\partial H}{\partial q_k}
= -\frac12 \sum_{i,j} p_i \, p_j \, \frac{\partial}{\partial q_k} 
M^{-1}_{ij} (\qvec) - \frac{\partial}{\partial q_k} U (\qvec). 
\end{equation}
Numerical integration of these equations is performed with 
the generalized leapfrog algorithm \cite{MEL3,SS,STLA,SIMHAM}, 
\begin{equation}
\label{GL1}
p_k^{n+\frac12} = p_k^n - \frac{h}{2} \left\{
\frac12 \sum_{i,j} p_i^{n+\frac12} \, p_j^{n+\frac12} \, 
\left(\frac{\partial M^{-1}_{ij}}{\partial q_k} \right)^n 
+ \left(\frac{\partial U}{\partial q_k}\right)^n\right\},
\end{equation}

\begin{equation}
\label{GL2}
q_k^{n+1} = q_k^n + \frac{h}{2} 
\sum_j p_j^{n+\frac12} \left\{ (M^{-1}_{kj})^{n+1} + (M^{-1}_{kj})^n \right\},
\end{equation}

\begin{equation}
\label{GL3}
p_k^{n+1} = p_k^{n+\frac12} - \frac{h}{2} \left\{
\frac12 \sum p_i^{n+\frac12} \, p_j^{n+\frac12} \, 
\left(\frac{\partial M^{-1}_{ij}}{\partial q_k} \right)^{n+1} 
+ \left(\frac{\partial U}{\partial q_k}\right)^{n+1}\right\},
\end{equation}
where $h$ is the time step and the superscript denotes the time-step number. 
While Eq.(\ref{GL3}) is an explicit formula, 
Eqs.(\ref{GL1}) and (\ref{GL2}) are implicit, 
i.e., these equations must be solved iteratively. 
Note, however, that we need to update only classical variables 
during the iterations, which is orders of magnitude faster 
than the evaluation of quantum mechanical forces ($ -\partial U / \partial \qvec$). 
Therefore, the computational cost of a single MTMD step is 
comparable to that of a conventional AIMD. 
Eqs.(\ref{GL1} - \ref{GL3}) also preserve the symplectic property, 
if the implicit ones are solved with sufficient accuracy \cite{SIMHAM}. 
When $M$ is indepedent of $\qvec$, the generalized leapfrog algorithm 
reduces to the velocity-Verlet method \cite{LIQ}. 

\section{Computational Details}
We carried out two AIMD simulations of liquid water to evaluate 
the performance of MTMD in real applications. 
The reference simulation (hereafter denoted by REF) 
was performed with 
$m_{\mbox{\tiny H}}=1.00794$ and $m_{\mbox{\tiny O}}=15.9994$, 
while the MTMD simulation was 
performed with $\qvec$-dependent masses, as will be explained below. 

Atomic forces were calculated within the density functional theory \cite{HK,KS}. 
We used the generalized gradient approximation 
in the Perdew-Burke-Ernzerhof form \cite{PBE}. 
The separable norm-conserving pseudopotentials were employed \cite{GTH,HGH}, 
and only the $\Gamma$ point was used to sample the Brillouin zone. 
The electronic orbitals were expanded by the 
finite-element basis functions \cite{FEM1,FEM2} with 
an average cutoff energy of 58 Ry, while the 
resolution was approximately doubled near the oxygen atoms 
by adaptation of the grids \cite{ACC}. 
Liquid water at an elevated temperature 
was modeled by 64 molecules in a cubic supercell of 
side 13.92 \AA  \cite{LIQW}, 
which was chosen to minimize the effect of 
nonergodic behavior observed at low temperatures \cite{VMKHS}. 
The equations of motion for the atoms were integrated with 
the generalized leapfrog algorithm 
using a time step of 20 a.u. (0.484 fs). 
Eqs.(\ref{GL1}) and (\ref{GL2}) were iterated 10 times 
to achieve full convergence in the MTMD run. 
Initial atomic velocities were chosen so that 
the total energies in the two simulations coincide. 
After equilibration, production runs of 15 ps were 
carried out in the microcanonical ensemble. 
The electronic states were quenched to the Born-Oppenheimer surface 
at each MD step 
with the limited-memory variant \cite{LINO,QNFEM} 
of the quasi-Newton method \cite{RCP}. 
The initial guesses were extrapolated from previous MD steps \cite{APJ}. 

Atomic masses used in the MTMD run were determined in the following manner. 
$V_{\mbox{\scriptsize H}}$ of Eq.(\ref{VHARM}) was first decomposed into a sum of 
contributions from each water molecule, 
\begin{equation}
V_{\mbox{\scriptsize H}} = \sum_{\mu} v_{\mu}. 
\end{equation}
The potential energy function $v_{\mu}$ for molecule $\mu$ can be written as 
\begin{equation}
\label{FFF}
v_{\mu} = \frac{k_1}{2} \left\{ (r_{1,\mu} - r_0)^2 + 
(r_{2,\mu} - r_0)^2 \right\} + 
\frac{k_2}{2} (\theta_{\mu} - \theta_0)^2, 
\end{equation}
where $r_1 = r$(OH$_1$), $r_2 = r$(OH$_2$), and $\theta = \angle$H$_1$OH$_2$. 
The parameters $k_1, k_2, r_0,$ and $\theta_0$ 
were determined by the force matching method \cite{FMM} 
using the equilibration part of the trajectories, as listed in Table \ref{FFPARAM}. 
Under this definition, $M$ is a block diagonal matrix, 
with each block being a square matrix of order 9. 
Therefore, inversion and diagonalization of $M$ 
can be carried out at negligible cost. 
In terms of the efficiency of phase space sampling, 
the value of $\epsilon$ introduced in Eq.(\ref{FILTER}) 
should be as small as possible, 
while too small a value may lead to instability. 
After some trial and error, 
$\epsilon = 0.1$ Ha/Bohr$^2$ proved to be a good compromise in the present system. 
The normalization factor $\alpha$ was adjusted so that 
the period of the fastest oscillation in MTMD agrees with 
that in REF, to make a fair comparison 
between the two simulations.

\section{Results}

\subsection{Numerical accuracy}
\begin{figure}
  \begin{center}
  \includegraphics[width=9cm]{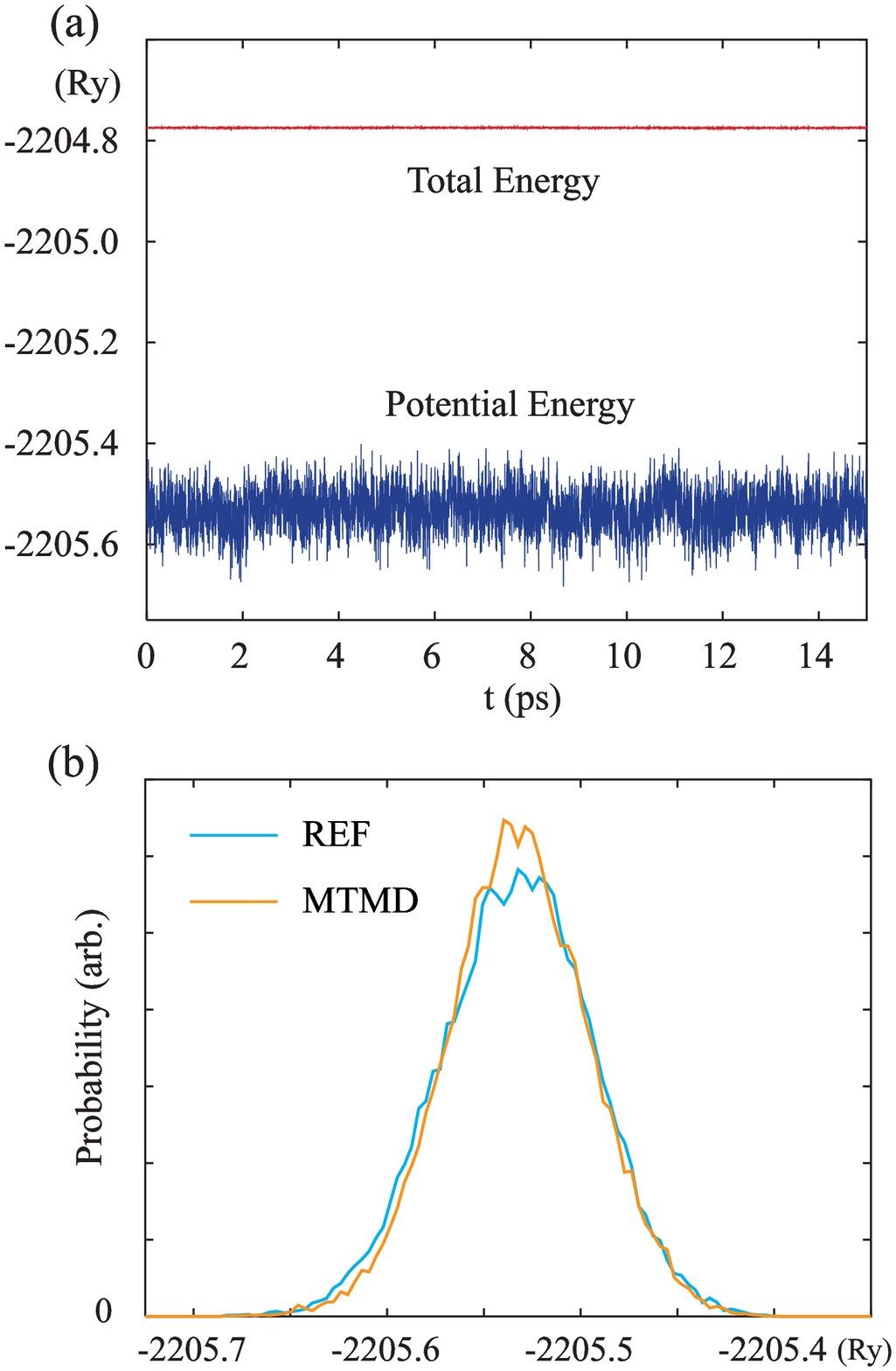}
  \end{center}
  \caption{Numerical accuracy of test calculations on liquid water. 
	   (a) Time evolution of total energy and potential energy in MTMD run. 
	   (b) Probability distributions of potential energies from REF and MTMD runs. 
	}
  \label{ENERGYFIG}
\end{figure}

We first show the time evolution of total energy and potential energy 
during the MTMD run in Fig.\ref{ENERGYFIG}(a), which reflects 
the accuracy of numerical integration. 
Conservation of the total energy in REF is also satisfactory. 
Average temperatures are 
415.74 K and 415.42 K for the REF and MTMD run, respectively. 
Probability distributions of potential energies 
are also compared in Fig.\ref{ENERGYFIG}(b). 
These results suggest that, within statistical errors, 
the two simulations are sampling the same region of phase space. 
We have also found that 
20 - 25 \% more computational effort is required for 
the calculation of interatomic forces in going from REF to MTMD. 
This is because the atoms move a longer distance at each MTMD step, 
as will be shown below, 
which makes the extrapolation of initial guesses less effective \cite{APJ}. 
A similar effect was observed in our previous work on CMD \cite{AIRAT}. 

\subsection{Structural properties}
\begin{figure}
  \begin{center}
  \includegraphics[width=9cm]{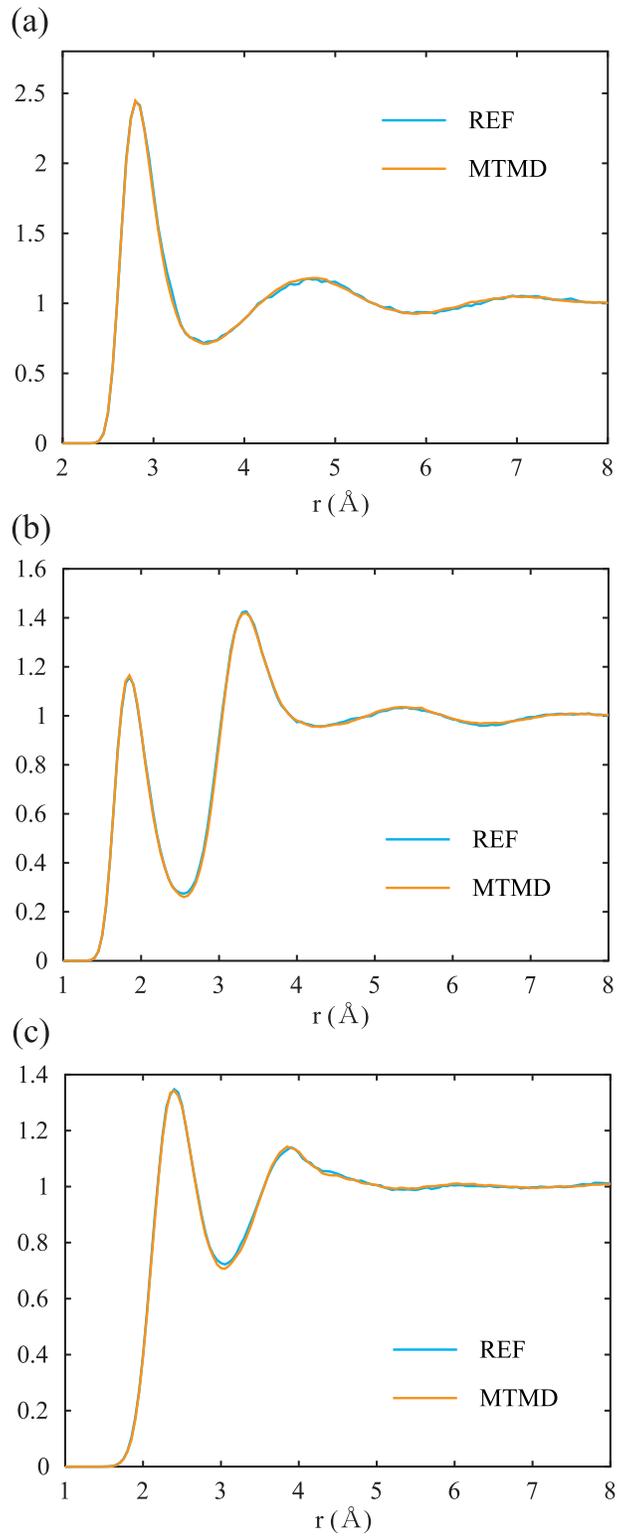}
  \end{center}
  \caption{(a) Oxygen-oxygen, (b) oxygen-hydrogen, and (c) hydrogen-hydrogen 
	radial distribution functions.} 
  \label{RDFFIG}
\end{figure}

\begin{figure}
  \begin{center}
  \includegraphics[width=9cm]{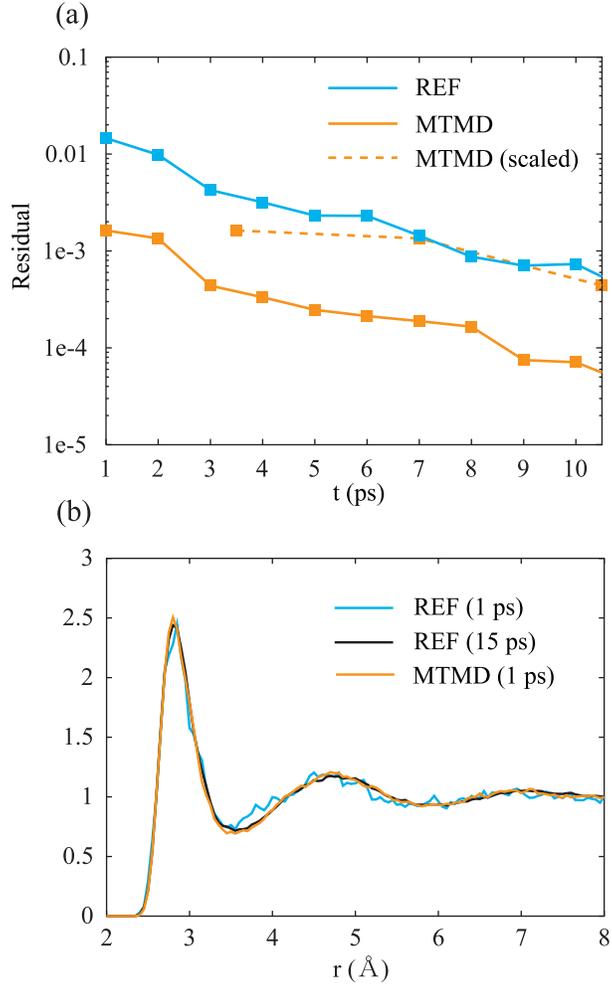}
  \end{center}
  \caption{Convergence of oxygen-oxygen RDFs ($g_{\mbox{\tiny OO}} (r,t)$) 
	with respect to simulation length $t$. 
	(a) Residual error $R(t)$ as a function of time. 
	 The scaled MTMD lines denote $R(t/3.5)$. 
 	(b) Comparison of $g_{\mbox{\tiny OO}} (r,t)$ at 
	 $t$ = 1 ps and $t = t_{\rm max}$.} 
  \label{CONVFIG}
\end{figure}

Average geometry of each molecule and the radial distribution functions (RDFs) 
have been calculated from the trajectories, 
as shown in Table \ref{AVEGEOM} and Fig.\ref{RDFFIG}, respectively. 
All results are in excellent agreement with each other, 
as expected from the theoretical analysis. 
We also compare the convergence of oxygen-oxygen RDFs ($\goo$) 
with respect to simulation length. 
The residual error in $\goo$ is defined by 
\begin{equation}
R(t) = \int_0^{r_{\rm max}} \left|\goo (r,t) - \goo (r,t_{\rm max})\right|^2 dr,
\end{equation}
where we set $r_{\rm max}$ = 7 \AA \ and $t_{\rm max}$ = 15 ps, 
and $\goo (r,t)$ denotes the oxygen-oxygen RDF, 
extracted from the trajectory in the range of $[0,t]$. 
As shown in Fig.\ref{CONVFIG}(a), the error for MTMD decays 
more than three times faster. 
The rapid convergence of MTMD is  more clearly seen in Fig.\ref{CONVFIG}(b), 
where the RDFs at $t=$ 1 ps 
are compared with that of $t=t_{\rm max}$. 
The residual errors for oxygen-hydrogen and hydrogen-hydrogen RDFs behave similarly, 
but are much smaller because there are twice as many hydrogen atoms 
in the simulation cell. 

\subsection{Dynamical properties}
\begin{figure}
  \begin{center}
  \includegraphics[width=9cm]{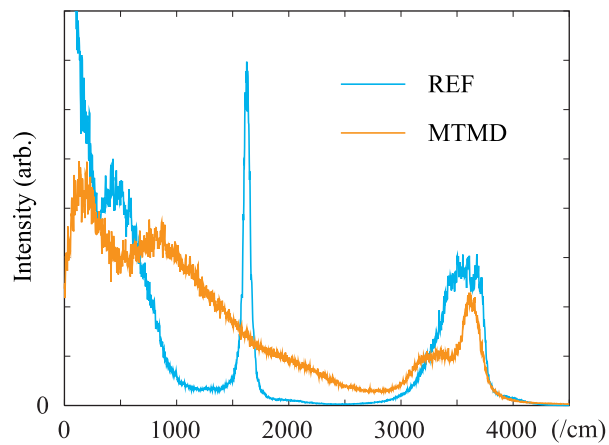}
  \end{center}
  \caption{Vibrational spectra of liquid water from REF and MTMD.}
  \label{VDOSFIG}
\end{figure}

\begin{figure}
  \begin{center}
  \includegraphics[width=9cm]{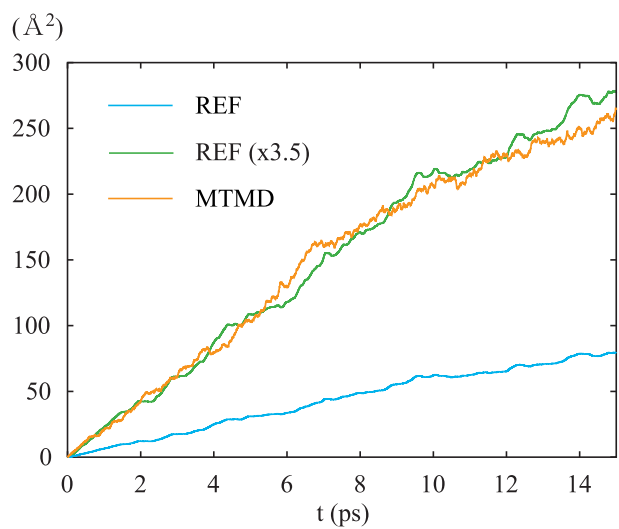}
  \end{center}
  \caption{Mean square displacements of oxygen atoms from REF and MTMD. 
	No average over time is taken.}
  \label{MSDFIG}
\end{figure}

The atomic velocities from the simulations 
have been used to calculate the vibrational spectra shown in Fig.\ref{VDOSFIG}. 
The first peak at 3400-3700 cm$^{-1}$ in REF corresponds to 
the O-H stretching mode, while the second peak at 1620-1630 cm$^{-1}$ 
is assigned to the H-O-H bending mode. 
The low frequency region below 1000 cm$^{-1}$ corresponds to the intermolecular modes. 
When going from REF to MTMD, 
the two peaks corresponding to the intramolecular modes 
merge into a single broad peak between 3200 - 3700 cm$^{-1}$. 
Furthermore, the intermolecular modes now extend up to 3000 cm$^{-1}$. 
Therefore, the gaps in the original spectra, which reduce the efficiency of simulations, 
are absent in the MTMD results. 
Note also that the highest frequency observed in MTMD 
is in good agreement with that in REF, 
which confirms that our choice of $\alpha$ is appropriate. 
This also means that 
MTMD and REF have the same accuracy, in the sense that the time step 
$\approx 1/20$ of the period of the fastest oscillation ($\approx 9$ fs) in both simulations. 
Finally, we compare the mean square displacements 
of oxygen atoms from the two simulations in Fig.\ref{MSDFIG}. 
In agreement with the convergence rate of RDFs, 
the mean square displacement from MTMD is found to grow 
$\approx 3.5$ times faster. 
If the aforementioned increase in computational cost is taken into account, 
the use of MTMD results in a net gain of 2.8 - 2.9 times. 
This gain is competitive with that of using a rigid water model \cite{CPRAT}, 
even though no fixed geometry approximation is required in MTMD.

\section{Discussion}

While ensemble averages of structural properties 
remain unchanged in MTMD, the same does not hold for 
dynamical properties like vibrational spectra and 
self-diffusion coefficients. 
This is because the trajectories satisfying 
the generalized equations of motion (Eqs.(\ref{NT1},\ref{NT2})) are unphysical, 
which can be viewed as a trade-off for enhanced sampling of the phase space. 
Therefore, MTMD should be used only in the equilibration phase, 
if dynamical properties are required. 
Nevertheless, there are several possible ways to 
recover the vibrational spectra 
of the original system within the harmonic approximation \cite{VIB3,VIB1,VIB4,VIB2}. 
Most of these methods rely on 
the covariance matrix of atomic fluctuations, 
\begin{equation}
\label{SGM}
\sigma_{ij} = \left<\,(q_i - \left<\,q_i\,\right>)(q_j - \left<\,q_j\,\right>)\,\right>,
\end{equation} 
which has the form of an ensemble average. 
Therefore, we can use the trajectory of MTMD or even Monte Carlo simulations 
to calculate Eq.(\ref{SGM}). 
In disordered systems, however, care must be taken in the choice of reference frame to 
take into account the effect of translation and rotation of each molecule. 
It is also worth noting that a novel algorithm 
based on the stochastic path-integral formalism \cite{SCALE} 
has recently been proposed, which allows us to recover the time correlation functions 
of the original system from the trajectory on a modified potential energy surface. 
A more empirical approach for estimating the self-diffusion coefficient 
from RDFs is also known for simple liquids \cite{BRETO}. 
The validity of these algorithms will be discussed in more detail elsewhere. 

As already mentioned in the Introduction, 
MTMD generally performs better than mass scaling, because 
the knowledge of the molecules can be fully exploited in MTMD. 
Nevertheless, mass scaling is still an attractive option, 
if chemical reactions can occur during the simulations \cite{SPES,SAW07,NAF1}. 
Moreover, at variance with MTMD, no programming effort is required. 
To the best of our knowledge, relatively little effort 
has been made \cite{MASS1} to optimize the atomic masses theoretically 
for finite temperature simulations, 
although several empirical studies exist \cite{FEEN,CAFES}. 
To this end, we first introduce the expression for the 
thermal average of the Hessian \cite{LIQ,VIB2}: 
\begin{equation}
\label{AVHESS}
\left< {\cal H}_{ij} \right> = 
\left< \frac{\partial^2 U}{\partial q_i \partial q_j} \right> = 
\frac{1}{k_{\rm B} T}
\left< 
\left( \frac{\partial U}{\partial q_i} \right)  
\left( \frac{\partial U}{\partial q_j} \right) 
\right>, 
\end{equation}
which was originally intended for vibrational analysis \cite{VIB2}. 
Here we propose to use Eq.(\ref{AVHESS}) to 
estimate the optimal masses $m_i^{\rm theory}$ by 
\begin{equation}
\label{MOPT}
m_i^{\rm theory} \propto \left< {\cal H}_{ii} \right> = \frac{1}{k_{\rm B} T}
\left< \left( \frac{\partial U}{\partial q_i} \right)^2 \right>,  
\end{equation}
where we take the average over all directions and atoms 
of the same element. Then, these masses may be used in conjunction with 
the original Hamiltonian of Eq.(\ref{OHAM}). 
Eq.(\ref{MOPT}) may be viewed as a special case of Eq.(\ref{MHEQ}). 
In the case of liquid water, we obtain a ratio of 2.0 for  
$m_{\mbox{\tiny O}}^{\rm theory} / m_{\mbox{\tiny H}}^{\rm theory}$, 
which is in close agreement with the empirical results of 
Feenstra {\it et al.} \cite{FEEN}. 
Theoretical prediction will play an important role in more complicated systems 
containing three or more elements 
where empirical optimization is prohibitive. 
Preliminary studies on the lithium borohydride (LiBH$_4$) system \cite{LIBH4} 
using the optimal masses are showing promising results. 
We also note that atoms of the same element may have different masses. 
For instance, hydrogen atoms forming O-H bonds may be made heavier 
than those forming C-H bonds, 
if these bonds remain stable throughout the simulations.

\section{Conclusion}

The importance of the choice of Hessian in {\it ab initio} geometry optimization 
is well known \cite{GOPT1,GOPTREV,GOPT4,GOPT6}. 
As we have shown in the present study, Hessian plays an equally important role 
in finite temperature simulations. 
MTMD will be the method of choice for nonreactive systems, 
if the primary interest of the simulations is 
the equilibrium properties \cite{METPAIR}. 
On the other hand, the mass scaling method is 
applicable to a wider class of problems, and will be 
particularly effective for systems with large differences in atomic masses, 
e.g., water/Pt interface \cite{OTN}. 
These approaches may also prove useful for global optimization problems \cite{ANNEAL}. 
It would also be interesting to combine MTMD with 
rare event methods such as hyperdynamics \cite{VTR} and metadynamics \cite{META}. 
Since these methods do not involve kinetic part of the Hamiltonian, 
the implementation would be straightforward.

\section*{Acknowledgements}
This work was supported by a KAKENHI grant (20038050). 
Numerical calculations were carried out on the T2K open supercomputer 
at the University of Tokyo.

\clearpage

\clearpage

\begin{table}[p]
\caption{Parameters of the force field for water molecules determined by 
the force matching method \cite{FMM}.}
\label{FFPARAM}
\begin{center}
\begin{tabular}{cccc}
\hline
\hline
$k_1$ (kcal/mol$\cdot$\AA$^2$) & $k_2$ (kcal/mol$\cdot$rad$^2$) & $r_0$ (\AA) & $\theta_0$ (deg) \\
\hline
846.87 & 74.33 & 0.9846 & 105.01 \\
\hline
\hline
\end{tabular}
\end{center}
\end{table}

\begin{table}[p]
\caption{Average structure of water molecules in liquid phase from the simulations.}
\label{AVEGEOM}
\begin{center}
\begin{tabular}{ccc}
\hline
\hline
     & $r$(OH) (\AA) &  $\angle$H$_1$OH$_2$ (deg) \\
\hline
REF    & 0.9856 $\pm$ 0.0308 & 104.80 $\pm$ 5.72   \\
MTMD  & 0.9857 $\pm$ 0.0312 & 104.84 $\pm$ 5.75   \\
\hline
\hline
\end{tabular}
\end{center}
\end{table}

\end{document}